\documentclass[letterpaper, 10 pt, conference]{ieeeconf}  

\IEEEoverridecommandlockouts                              

\usepackage{comment}
\usepackage[dvipdfmx]{graphicx}
\usepackage{enumerate}
\usepackage{fancyhdr}

\title{\LARGE \bf
Presenting Static Friction Sensation \\ at Stick-slip Transition using Pseudo-haptic Effect}

\author{Yusuke Ujitoko$^{1}$, Yuki Ban$^{2}$ and Koichi Hirota$^{3}$
\thanks{$^{1}$ Yusuke Ujitoko is with Research \& Development Group, Hitachi, Ltd., Yokohama, Japan and is a graduate student of the University of Electro-Communications.
        {\tt\small yusuke.ujitoko.uz@hitachi.com}}%
\thanks{$^{2}$ Yuki Ban is with the Mechanical Engineering Department, the University of Tokyo, Chiba, Japan.
        {\tt\small ban@edu.k.u-tokyo.ac.jp}}%
\thanks{$^{3}$ Koichi Hirota is with the Graduate School of Information Systems, The University of Electro-Communications, Tokyo, Japan.
        {\tt\small hirota@vogue.is.uec.ac.jp}}%
}

\begin{document}

\maketitle

\thispagestyle{fancy}
\fancyhf{}
\lhead{ {\tiny 2019 IEEE.  Personal use of this material is permitted.  Permission from IEEE must be obtained for all other uses, in any current or future media, including reprinting/republishing this material for advertising or promotional purposes, creating new collective works, for resale or redistribution to servers or lists, or reuse of any copyrighted component of this work in other works.\\}}
\renewcommand{\headrulewidth}{0pt}


\begin{abstract}
Previous studies have aimed at creating a simple hardware implementation of surface friction display.
In this study, we propose a new method for presenting static frictional sensation using the pseudo-haptic effect as a first attempt, which is the simplest implementation of presenting static friction sensation.
We focus on the stick-slip phenomenon while users explore surfaces with an input device, such as a stylus.
During the stick phase, we present users with pseudo-haptic feedback that represents static friction on the surface.
In our method, users watch a virtual contact point become stuck at the contact point on screen while users freely move the input device.
We hypothesize that the perceived probability and intensity of static friction sensation can be controlled by changing the static friction coefficient as a visual parameter.
User studies were conducted, and results show the threshold value over which users felt the pseudo-haptic static friction sensation at 90\% probability.
The results also show that the perceived intensity of the sensation changed with respect to the static friction coefficient.
The maximum intensity change was 23\%.
These results confirm the hypothesis and show that our method is a promising option for presenting static friction sensation.
\end{abstract}

\section{Introduction}


Friction displays that present tangential forces over a users'{} fingers have been developed in other studies.
The studies fell into two categories.
First, force displays represent the actual frictional force on the contact surface \cite{Hayward}.
Second, the frictional characteristics of the contact surface were changed \cite{tPad}, \cite{Nara:2001:SAW:616073.618864}.
Studies in both categories could effectively present a frictional force to users.
However, in terms of accuracy, power consumption, and size, mechanical add-ons for devices (e.g., mobile devices) are often impractical.
In such cases, a method for presenting frictional feedback without additional equipment is preferable.

On the other hand, some studies increasingly focus on pseudo-haptics.
Pseudo-haptics is a cross-modal effect between visual and haptic senses \cite{Lecuyer2009}.
The pseudo-haptic effect indicates the haptic perception evoked by vision.
A sensation is produced by an appropriate sensory inconsistency between the physical motion of the body and the observed motion of a virtual pointer.
For example, when a pointer decelerates in a standard desktop environment with a mouse , users feel a kinetic frictional force without any haptic actuator \cite{Lecuyer2000}.

While kinetic friction sensation using pseudo-haptics was presented in some papers \cite{Lecuyer2000, Narumi2017, Virtual_string}, static friction sensation was not addressed.
However, the methods in these studies are ineffective for rendering the fricitonal properties of materials with the same kinetic friction coefficients but different static friction coefficients.
Static friction sensation should be presented in order to allow users to recognize and discriminate various material surfaces with a diverse range of static coefficients.

\begin{figure}[t]
   \centering
   \includegraphics[width=3.3in]{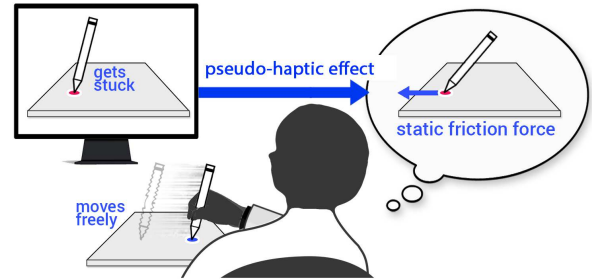}
   \caption{We propose a method for presenting static friction sensation using pseudo-haptics. During the stick phase in the stick-slip phenomenon, the virtual pointer sticks to a surface and users feel pseudo-haptic static friction.}
   \label{fig_main_image}
\end{figure}

In this study, we propose a method for presenting static friction sensation using the pseudo-haptic effect.
We focused on the stick-slip phenomenon while users explore surfaces with an input device, such as a stylus.
During the stick phase, users were presented with pseudo-haptic friction sensation, which represents the frictional properties of a material (see Fig.\ref{fig_main_image}).
However, if we implement the concept in a straightforward way, the visualized contact point becomes stuck and appears to have no relation to a user's input.
As a result, the sense of agency over the point would be lost and it would prevent the induction of pseudo-haptics.
Thus, we applied an additional virtual string technique \cite{Virtual_string} to maintain the sense of agency. Details are described in Section 3.
We hypothesize that we can control the perceived probability and intensity of static friction sensation by changing the visual parameters.
We conducted user studies to test this hypothesis.



\section{Related work}
\subsection{Real Frictional Feedback}

Presenting frictional feedback ordinarily requires presenting a tangential force, and presenting it with a mechanical interface has been widely researched
\cite{Hayward,tPad,Nara:2001:SAW:616073.618864,alternative_konyo}. 
Although these approaches can elicit frictional feedback, applying them to handheld devices is often impractical because additional electro-mechanical components are required.

%

\subsection{Pseudo-haptic Feedback}

Recent seminal work focused on pseudo-haptics, which made full use of the cross-modal effect to render haptic sensation.
Pseudo-haptic sensation occurs when physical body motion differs from the observed motion of a virtual pointer on a screen \cite{Lecuyer2009}.
When a user believes that the pointer moves according to the movement of their body, changes in the movement of the pointer are regarded as changes in the haptic sensation, such as a force on the hands, evoking a pseudo-haptic sensation.

Regarding texture perception, attempts were made in prior research to generate texture perception using pseudo-haptic effects. 
These studies aimed to provide perception of macro roughness \cite{Lecuyer2004}, fine roughness \cite{Ujitoko_pseudo_roughness}, friction \cite{Lecuyer2000, Ujitoko2015, Narumi2017, Virtual_string}, and stiffness \cite{Argelaguet:2013:EIP:2506206.2501599, Hachisu2011}.


Studies that focused on friction sensation presented kinetic friction sensation while a virtual pointer slips on a surface.
For example, users were allowed to feel the kinetic friction by simply using variations in the motion of the pointer without any haptic device in \cite{Lecuyer2004}.
A similar technique was applied in the touchscreen environments in \cite{Narumi2017}.
The use of a virtual string that showed a connection between a finger and the pointer on the touchscreens was proposed in \cite{Virtual_string}, which maintained a sense of agency over the pointer. We adopted the additional virtual string \cite{Virtual_string} in this study.

While kinetic friction sensation was addressed in some studies, static friction sensation was not addressed.
It is necessary to present static friction sensation in order to allow users to recognize and discriminate material surfaces which have a diverse range of static friction coefficients.

\section{Concept and Implementation}

\subsection{Concept and Hypotheses}

The objective of this study is to present static friction sensation using the pseudo-haptic effect.
We focus on the stick-slip phenomenon between an input device and surfaces.
The stick-slip phenomenon is generally a dynamic cyclic process where two contacting surfaces oscillate between a stick phase and a slip phase.
During the stick phase, the two surfaces are not in motion and are held in place by static friction.
During the slip phase, there is finite relative motion where kinetic friction acts to retard this movement.

In the real world, when we attempt to slide the stuck pen on a surface whose static friction coefficient is large, and a progressively larger tangential force would be applied to the pen.
When the applied tangential force is greater than the maximum force of static friction, the surfaces start to slide.

\begin{figure}[h]
   \centering
   \includegraphics[width=2.9in]{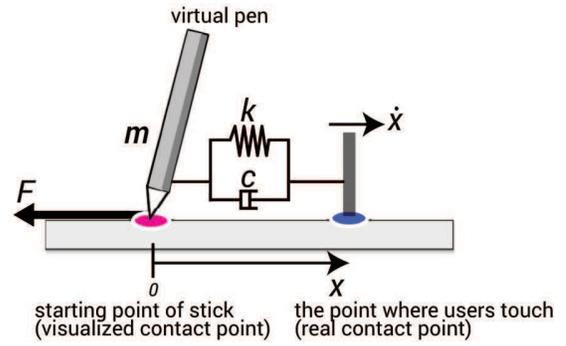}
   \caption{Model of stick-slip phenomenon.}
   \label{fig_stick_slip_model}
\end{figure}

Thus, we hypothesize that users would feel a pseudo-haptic static friction sensation because due to the visuo-haptic sensory inconsistency if a user watches the virtual point of touch getting stuck at the contact point while the user slides a real input device.
Fig.\ref{fig_main_image} illustrates this concept.
Fig. \ref{fig_stick_slip_model} shows the visualized contact point and the real contact point. They are modeled by Coulomb's model, which will be described later.

Fig.\ref{fig_visuo-haptic} shows the visuo-haptic sensory inconsistency during stick phase.
The figure assumes a case where the input device moves at a constant speed (blue line) and the visualized pointer sticks and slips (red line).
The visuo-haptic sensory inconsistency increases during the stick phase.

\begin{figure}[h]
   \centering
   \includegraphics[width=2.8in]{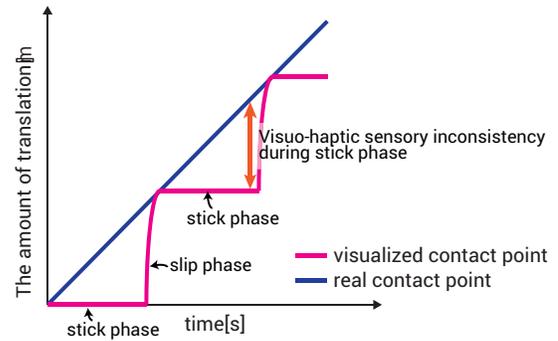}
   \caption{Translation the between visualized contact point and the real contact point. Visuo-haptic inconsistency during the stick phase would generate a pseudo-haptic static friction sensation.}
   \label{fig_visuo-haptic}
\end{figure}

The maximum length of the visuo-haptic sensory inconsistency is defined by the stick-slip model's static coefficient parameter if other parameters are fixed (described in the next section).
Previous studies \cite{Lecuyer2000, Virtual_string} show that the level of visuo-haptic sensory inconsistency affect the intensity of pseudo-haptics.
Thus, we also hypothesize that the configuration of the static friction coefficient would affect the pseudo-haptic friction sensation.
For example, a larger static friction coefficient makes the virtual pointer stickier.
In contrast, a smaller static friction coefficient makes the virtual pointer less sticky.

\subsection{Problem}

However, there emerges a problem of losing a sense of agency.
This arises because the input movement apparently does not affect the visualized pointer during the stick phase.
As a result, the sense of agency over the point would be lost and it would prevent induction of pseudo-haptics.

In order to prevent this problem and maintain the sense of agency over the pointer, we added a virtual string \cite{Virtual_string}.
The virtual string extends from the contact point as the user moves the input device.
The virtual string reflects the user's input, thus the sense of agency should be maintained.
Implementation of a virtual string is described in the next section.

\begin{figure}[h]
   \centering
   \includegraphics[width=3.3in]{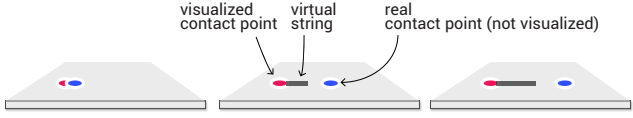}
   \caption{The virtual string elongates as the user moves the input device while the visualized virtual contact point remains stuck.}
   \label{fig_visuo-haptic}
\end{figure}


The hypotheses tested in this study are summarized as follows:

\begin{itemize}
  \item {\bf H1}\,\, If users watch the pointer visually stick to a point while the user freely slides the input device, they believe that the static friction on the surface is larger. Also, the presence of a virtual string would affect the perceived probability of static friction.
  \item {\bf H2}\,\, As the static coefficient setting becomes larger, the intensity of the sensation would be larger.
\end{itemize}

To test the hypotheses, we conducted two user studies.
User studies 1 and 2 tested H1 and, respectively.

\subsection{Implementation Overview}

\begin{figure}[h]
   \centering
   \includegraphics[width=3.1in]{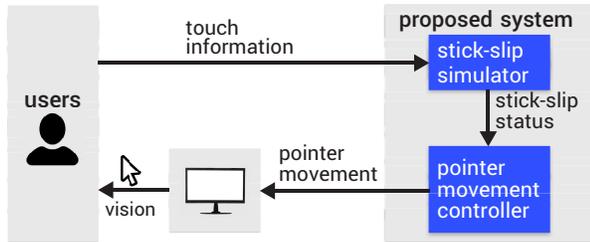}
   \caption{Data flow during interaction between a user and the system.
A user inputs touch information to the system. The proposed system provides the user with distorted pointer movement as visual information.}
   \label{fig_data_flow}
\end{figure}

We implemented a system that implements the concept and used it to conduct studies with users.
Data flow during interaction between the user and the system is illustrated in Fig.\ref{fig_data_flow}.

We assume use cases where a user explores a surface using a mouse or a touch pen device.
A touch pen was used in this study.
A user holds the pen and moves it on a touchpad.
Touch information, such as touch timing and position on the touchpad, is transmitted to the proposed system when a user moves the pen.
The system simulates a stick-slip phenomenon based on touch information and visualizes the pointer.
The system has two key modules: a stick-slip simulator and a pointer movement controller.

The stick-slip simulator plays a role in simulating a physical interaction between a pen and a surface in the virtual world.
The simulator updates the stick/slip phase, position, velocity, amnd acceleration of the pen.
The simulator informs the other module of the stick-slip status.

The pointer movement controller plays a role in visualizing the position of the virtual pointer, which is based on the stick-slip status.
An ordinary system does not require modifying the pointer position and it returns the position as a user moves the pen.
In contrast, our system manipulates the pointer position based on the stick-slip simulator.
A user feels pseudo-haptic static friction by watching the visualized pointer.
We described how these modules are implemented in the rest of this section.

\subsection{Stick-slip Simulator}


We used Coulomb's stick-slip model \cite{Nakano2006}, which focuses on representing the stick-slip phenomenon with a simple analytical model.
The friction model is composed of a virtual pen (mass $m$), a linear spring (stiffness $k$), and a viscous damper (damping coefficient $c$), as shown in Fig.\ref{fig_stick_slip_model}.
The virtual pen contacts a surface with a normal load $mg$, and a friction force $F$ is applied to the pen.
The user touches the point $x$.
The point $x$ and the virtual pen are connected with the spring and damper.
%

During slip, the motion of the virtual pen is described as follows:

\begin{eqnarray}
\label{eq:F_k}
  m\ddot{x} + c\dot{x} + kx &= F_k,  &if\,\, \dot{x} < 0,\\
  m\ddot{x} + c\dot{x} + kx &= -F_k, &if\,\, \dot{x} > 0
\end{eqnarray}

where $x$ denotes the position of the pen and the natural length of spring is $0$.
The kinetic friction force is $F_k = \mu_k \cdot mg$.
During slip, slip-to-stick transitions occur when the velocity of the virtual pen $\dot{x}$ equals $0$.

During stick, $\ddot{x}=\dot{x}=0$ and stick-to-slip transitions occurred when $kx > F_{smax}$,
where $F_{smax} = \mu_s \cdot mg$.
When a stick-to-slip transition occurs, the virtual pointer should instantly translate to the point where user is touching.
In other words, setting the critical damping parameter is preferable instead of allowing damped oscillation.
Thus, we define the parameter of the mass $m$ based on the damping coefficient $c$ and stiffness $k$ such that they satisfy critical dumping.

\subsection{Pointer movement controller}


During stick, a user is presented with pseudo-haptic feedback, which represents a sensation of friction.
The pointer movement controller simply visualizes the virtual pointer based on the stick-slip simulation.
During stick, the pointer does not move and a user would feel pseudo-haptic friction while the user moves the pen.

In addition, we added a virtual string \cite{Virtual_string} that shows the connection between the point of the virtual pen and the point where the user touches the pen.
It has a role in maintaining the sense of agency over the pointer, even if the point sticks.
According to the findings of \cite{Virtual_string}, the length of the string should change based on the stick friction, which equals the virtual force applied to the string.
We conducted informal studies and optimized the visualization of the string.
As a result, we defined the length $l$ of the string heuristically as $l = C_{l} \sqrt{F_s}$.
An informal study and the results show that the presence of the virtual string affects pseudo-haptics.


\section{User study 1}

User study 1 was conducted to test hypothesis H1.
Specifically, we tested the probability that participants feel larger surface static friction when they watch a virtual pointer sticking to the contact point.
In addition, we tested whether or not the perceived probability can be influenced by the presence of a virtual string.
Thus, we conducted an experiment with and without a virtual string.
In other words, we performed two within-participant experiments and compared six different visual conditions.

There were 10 participants (eight males and two females) aged from 22 to 25.
All participants were right-handed.
They were screened to determine that they were not depressed or tired because perception can be affected by physical or emotional stress.
The University of Tokyo Ethics Committee approved the experiments presented in this paper and written informed consent was obtained from all participants in the studies presented here and in the next section.

\subsection{Experimental system}

\begin{figure}[t]
   \centering
   \includegraphics[width=3.3in]{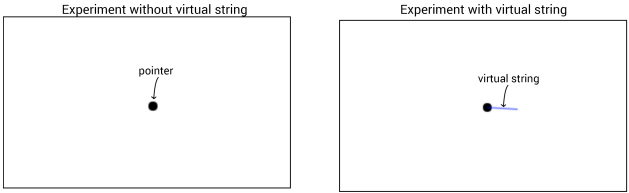}
   \caption{Experimental window with and without a virtual string.}
   \label{fig_ex1_window}
\end{figure}

\begin{figure}[t]
   \centering
   \includegraphics[width=2.5in]{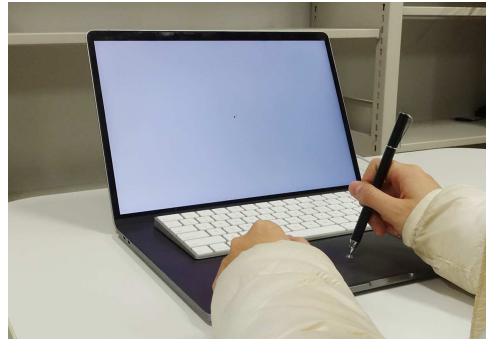}
   \caption{Experimental system.}
   \label{fig_experimental_scene}
\end{figure}

Participants had to move the pen from left to right on the touchpad while watching a virtual pointer visualized on the screen.
The experimental window is simple and is shown in Fig.\ref{fig_ex1_window}.
The pointer colored in black was visualized at the center and participants' task is to move the pointer 70 pixels distance from the center.
The pointer refers to the contact point, although the position of the pointer was controlled by the pointer movement controller.
The virtual string extended from the pointer as users move the pen on the touchpad.
In contrast, the string was not visualized in the condition without a string.
Our experimental system was composed of a touchpad and display (2880 pixel x 1800 pixel, retinal display) on the laptop PC (Apple Inc., MacBookPro) shown in Fig. \ref{fig_experimental_scene}.
Participants wore noise canceling headphones and heard white-noise during the experiment to suppress background noise.

The following parameters were used in this study: $g = 9.8 m/s^2$, $k = 0.1 N/m$, $\mu_k = 0.1$, and $C_l = 2000$.
The stick-slip simulator operated at 100 Hz.

\subsection{Task design}

The experiment was designed by following a just noticeable differences (JND) methodology \cite{gescheider1985psychophysics}.
The JND experiment requires participants to choose a reference value for the standard stimulus (reference target) and compare this with a comparison target.
For each trial, participants performed the standard stimulus and the comparison stimulus sequentially, and they stated which stimulus made the surface feel more frictional.
The static friction coefficient was different between standard stimuli and comparison stimuli.
The static friction coefficient $\mu_s$ of a standard stimulus was always 0, thus there was no stick phase.
On the other hand, $\mu_s$ for the comparison stimulus was set to 0.0, 0.2, 0.4, 0.6, 0.8, or 1.0. 

The pointer was positioned at the center of the window when stimulus began.
The pointer was in the stick phase when $\mu_s$ was greater than 0.
After participants moved the pen a certain distance (which depends on the $\mu_s$), the pointer moved into the slip phase and began to move.
When the pointer moved 70 pixels from the center position, the experiment moved into the next stimulus or evaluation phase.
After participants finished moving the virtual pointer with both stimuli, they stated which stimulus made the surface feel more frictional.
They tapped one of two answer buttons visualized on the screen.
We told them to select one of two buttons randomly if they thought that it was difficult to judge which was rougher.

There were six visual conditions for comparison stimulus.
Participants performed the experiment with each condition 10 times.
Thus, each participant conducted 60 trials with and without the virtual string.
These trials were ordered randomly and were counterbalanced across participants.

\subsection{Results}

\begin{figure}[h]
   \centering
   \includegraphics[width=3.3in]{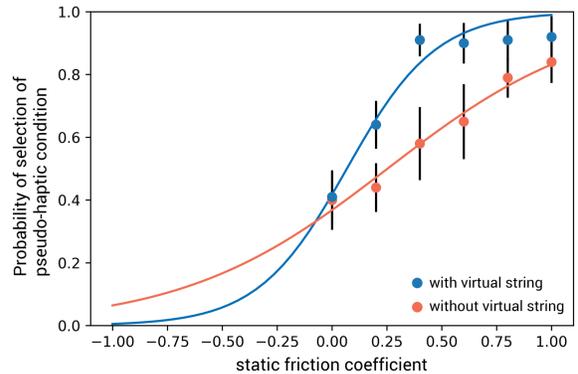}
   \caption{Probability that participants felt the comparison stimulus provided greater frictional than the standard stimulus as a function of static friction coefficient. The standard error for each plot is also shown.}
   \label{fig_ex1_result}
\end{figure}

Fig.\ref{fig_ex1_result} shows the experimental results with and without the virtual string.
This shows the averaged probability that participants answered that the comparison stimulus felt more frictional than the standard stimulus.
We calculated the psychometric function for the condition with/without the string to analyze the minimum noticeable difference.
The JND results provide insight into the minimum difference in the friction sensation that can be distinguished.
The perceived probability curve was obtained fitting the psychometric curve to the data $f(x) = \frac{1}{1+\exp(-A \cdot (x-B))}$. Calculated $A$ and $B$ values for the experiment with the string were $0.068$ and $4.9$, respectively. $A$ and $B$ for the experiment without the string were $0.25$ and $2.1$, respectively. JND is equal to the static friction coefficient $\mu_s$ at the 75\% point on this curve minus the point of subjective equality (PSE). The JND value of $\mu_s$ with the virtual string was $0.29$ and the value without the string was $0.77$.

\subsection{Discussion}

Based on the results in Fig.\ref{fig_ex1_result}, participants felt the pseudo-haptic static friction when the static friction coefficient exceeded a certain threshold.
This proved the concept of this study described in the previous section.

We could present the pseudo-haptics under the condition with a string rather than without a string.
The calculated JND value provides evidence for the effectiveness of the use of a string.
When $\mu_s$ = 0.29, users felt the pseudo-haptics with a string, although they did not without a string.
Collection of the free comments after the user study showed
that four participants out of ten reported that a difference in feeling with or without string.
They said that they found the task difficult without the string.
When the static coefficient ranges from 0.4 to 1.0, participants felt the pseudo-haptic static friction at greater than 90\% under the condition with string.
The effectiveness is assumed to originate from the sense of agency that is caused by the virtual string.

\section{User study 2}

We performed a within-participant study in User Study 2 comparing seven different visual conditions.

User Study 2 was conducted to test hypothesis H2.
In other words, we tested whether the intensity of the sensation would be larger as the static coefficient setting increases.
This study focused on the condition with the string because users felt pseudo-haptics robustly under that condition.

There were 10 participants (eight males and two females) aged from 22 to 25.
All participants were right-handed.
They were screened to determine that they were not depressed or tired because perception can be affected by physical or emotional stress.

\subsection{Experimental system}
The experimental system was the same as that used in User Study 1.
The system was composed of a touchpad, display, and pen.

\subsection{Task design}

The task design follows that used in User Study 1.
This study used a within-participants design.
Participants moved a pen from left to right or from right to left while watching the virtual pointer.
They initially used a standard stimulus and subsequently used a comparison stimulus.
According to the results of User Study 1, a user felt pseudo-haptic static friction with virtual string when the static friction coefficient ranged from 0.4 to 1.0.
Based on this result, we set the static friction coefficient of the standard stimulus to 0.7.
The static friction coefficient of the comparison stimulus was set to 0.4, 0.5, 0.6, 0.7, 0.8, 0.9, or 1.0.

The procedure for one trial in the task was nearly the same as that used in User Study 1.
The only difference was that participants evaluated the ratio of the perceived comparison stimulus intensity with to the standard stimulus intensity.
The task design resembles that used in previous research \cite{10.1007/978-3-540-69057-3_79} in terms of evaluation of perceived intensity between standard and comparison stimuli.
Participants pushed a button from "decrease (-0.10)", "slight decrease (-0.05)", "slightest decrease (-0.01)", "slightest increase (+0.01)", "slight increase (+0.05)", or "increase (+0.10)" on the screen to assign numbers to quantify the subjective intensity ratio.
The initial intensity ratio configuration was set to 1.0.
There was no time limit for adjustment and participants could push buttons repeatedly.

There were seven visual conditions for comparison stimulus.
Participants repeated the experiment in each condition 5 times.
Thus, each participant conducted 35 trials.
These factors were presented in a random order and were counterbalanced across participants.

\subsection{Results}

\begin{figure}[h]
   \centering
   \includegraphics[width=3.3in]{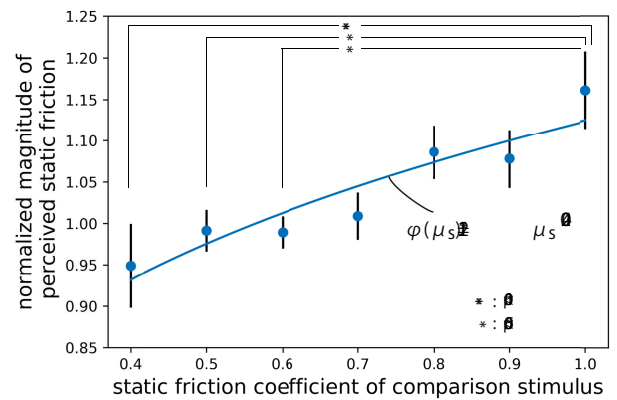}
   \caption{The ratio of the subjective intensity between the comparison and standard stimuli.
   The standard error for each plot and the fitting curve to Steven's power function are also shown.}
   \label{fig_ex2_result}
\end{figure}

Fig.\ref{fig_ex2_result} shows the results.
The horizontal axis shows the static friction coefficient of the comparison stimulus, and the vertical axis shows the ratio of the pseudo-haptic intensity of the comparison stimulus to that of the standard stimulus.

To determine whether the participants felt more intense pseudo-haptic static friction as the static friction coefficient $\mu_s$ increased, we performed a one-way repeated ANOVA with factors of $\mu_s$ values ($\mu_s = 0.4, 0.5, 0.6, 0.7, 0.8, 0.9, or 1.0$) on the ratio.
We conducted a Shapiro-Wilk test to check the normality and a Mauchly's test to check the sphericity criteria in advance of the ANOVA test.
According to the ANOVA results, the static friction coefficient significantly affects the perceived intensity ($F(6, 54) = 4.22, p = 0.0012$).

We applied Tukey comparisons for all post-hoc comparisons for the amplitude experiment.
As a result, there was a significant difference between the conditions where $\mu_s$ from the comparison stimulus ranged from 0.4 to 1.0 ($p \verb|<| 0.01$), 0.5 to 1.0 ($p \verb|<| 0.05$), and 0.6 to 1.0 ($p \verb|<| 0.05$).

The task in this user study followed magnitude estimation.
The data was fit to Steven's power function $\phi(\mu_s) = k \mu_s^{\beta}$.
The calculated values of $k$ and $\beta$ were 1.12 and 0.204, respectively.

\subsection{Discussions}

ANOVA and Tukey comparisons (Fig.\ref{fig_ex2_result}) show that the static friction coefficient significantly affects the perceived pseudo-haptic intensity.
This shows that participants felt stronger friction as the static friction coefficient became larger.
In other words, hypothesis H2 was confirmed.
The obtained parameter from Steven's power function also confirmed H2.
Based on this user study, we can modulate the perceived intensity of static friction by changing the value of the static friction coefficient.

The ratio was 0.94 when $\mu_s$ = 0.4, and the ratio was 1.16 when $\mu_s$ = 1.
The ratio of these two ratios was 1.23.
This shows that we can change the perceived intensity of static friction at least 23\% by changing $\mu_s$ from 0.4 to 1.

The results from user studies show that the proposed method can present static friction using a pseudo-haptic effect with high probability.
In addition, the perceived intensity changes with parameter changes.
Our method only requires visual information and is applicable to cases where other methods requiring mechanical add-ons cannot be applied.

Some things remain unclear for now.
For example, we defined the extension function of the virtual string heuristically in this study, and it is unclear whether the pseudo-haptic effect changes when a different function is used.
It is also unclear whether our method is effective when the static friction coefficient is greater than 1.0.
We leave these interesting topics for a future study.

\section{CONCLUSIONS}

A method for presenting static friction using the pseudo-haptic effect is proposed in this paper.
The user studies yielded the following findigs:

\begin{itemize}
  \item When users watch the pointer visually stick to a particular point while freely sliding the input device, they believe that the surface static friction is larger. The effect occurs with greater than 90\% probability when the visual parameters exceeded a threshold value. Visualizing a virtual string \cite{Virtual_string} increased the probability.
  \item The perceived intensity of the sensation would be larger as the static coefficient setting increases. The maximum intensity change confirmed in the user study was 23\%.
\end{itemize}

These results suggests that our method is helpful for presenting static friction sensation with high probability.
Our method is simple and can be implemented with current off-the-shelf mobile devices without any haptic actuator.

\addtolength{\textheight}{-12cm}   





%


\bibliographystyle{IEEEtran}
\bibliography{whc_pseudoFriction}

\end{document}